\documentclass[11pt,amssymb,nofootinbib]{revtex4}
\usepackage[latin9]{inputenc}
\usepackage{textcomp}
\usepackage{amsmath}
\usepackage{amssymb}
\usepackage{esint}
\usepackage{graphicx}
\usepackage{epstopdf}
\usepackage[english]{babel}

\usepackage[latin9]{inputenc}
\usepackage{textcomp}
\usepackage{amsmath}
\usepackage{amssymb}
\usepackage{esint}
\usepackage{graphicx}
\usepackage{epstopdf}
\usepackage[english]{babel}
\usepackage{verbatim}
\usepackage{bm}
\usepackage{color,graphicx}
\usepackage{tabularx}
\usepackage{times}

\usepackage{amsthm}\usepackage{amscd}\usepackage{bm}\usepackage{bbm}

\usepackage{feyn}

\makeatother

\begin{document}

\title{Novel CSL bounds from the noise-induced radiation emission from atoms}

\author{Sandro Donadi}
\affiliation{Istituto Nazionale di Fisica Nucleare, Trieste Section, Via Valerio 2, 34127 Trieste, Italy}

\author{Kristian Piscicchia}
\email{kristian.piscicchia@cref.it}
\affiliation{Centro Ricerche Enrico Fermi - Museo Storico della Fisica e Centro Studi e Ricerche ``Enrico Fermi'', Piazza del Viminale 1, 00184 Rome, Italy}
\affiliation{INFN, Laboratori Nazionali di Frascati, Via Enrico Fermi 54, 00044 Frascati, Italy.}

\author{Raffaele Del Grande}
\affiliation{Physik Department E62, Technische Universit\"at M\"unchen, James-Franck-Stra\ss e 1, 85748 Garching bei M\"unchen, Germany.}

\author{Catalina Curceanu}
\email{catalina.curceanu@lnf.infn.it}
\affiliation{INFN, Laboratori Nazionali di Frascati, Via Enrico Fermi 54, 00044 Frascati, Italy.}

\author{Matthias Laubenstein}
\affiliation{INFN, Laboratori Nazionali del Gran Sasso, Via G. Acitelli 22, 67100 Assergi, Italy.}

\author{Angelo Bassi}
\affiliation{Department of Physics, University of Trieste, Strada Costiera 11, 34151 Trieste, Italy\looseness=-1}
\affiliation{Istituto Nazionale di Fisica Nucleare, Trieste Section, Via Valerio 2, 34127 Trieste, Italy.}

\begin{abstract}
We study spontaneous radiation emission from matter, as predicted by the Continuous Spontaneous Localization (CSL) collapse model. We show that, in an appropriate range of energies of the emitted radiation, the largest contribution comes from the atomic nuclei. Specifically, we show that in the energy range $E\sim 10\,-\,10^{5}$ keV the contribution to the radiation emission from the atomic nuclei grows quadratically with the atomic number of the atom, overtaking the contribution from the electrons, which grows only linearly. This theoretical prediction is then compared with the data from a dedicated experiment performed at the extremely low background environment of the Gran Sasso underground National Laboratory, where the radiation emitted form a sample of Germanium  was measured. 
As a result, we obtain the strongest bounds on the CSL parameters for $r_C\leq 10^{-6}$ m, improving the previous ones by more than an order of magnitude. 
\end{abstract}

\maketitle

\section{Introduction} 
The superposition principle is one of the cornerstones of Quantum Mechanics. Microscopic systems like electrons, atoms, etc. can be in a superposition of states corresponding to different positions and this is at the root of many phenomena which cannot be explained within classical physics. The superposition principle follows from the linearity of the Schr\"odinger equation. If one assumes this linearity to be universal, the theory leads to puzzling predictions, as the existence of macroscopic superpositions like the Sch\"odinger's dead and alive cat~\cite{gatto}. To avoid this problem it has been suggested that the quantum linear evolution breaks down beyond a certain scale~\cite{leggett, weinberg, bell, Pen1996}.

The departure from the linear evolution for the state vector has been studied in great detail in collapse models: these are phenomenological models which assume that the collapse of the wave function is accounted for by a non-linear interaction with an external classical noise~\cite{grw,csl,rep1,rep2}. The interaction gives extremely small deviations from the Sch\"odinger's dynamics for microscopic systems; when, instead, a macroscopic system is considered, the collapse takes over, with probabilities in agreement with the Born rule. 

In addition to collapsing the wave function in space, the interaction with the classical noise induces a diffusion in space, resulting in a Brownian-like motion. 
For a system made of charged particles, this Brownian-like diffusion forces the particles to emit radiation. Since this phenomenon is not predicted by standard quantum mechanics, the noise-induced radiation emission can be used to set bounds on the collapse parameters.
This idea was recently used in the context of gravity-related wave function collapse~\cite{Diosi1, Diosi2, Pen1996}, leading to strong bounds on these models~\cite{NaturePhys}.

The noise-induced radiation emission in collapse models was extensively studied~\cite{fu,ar,bd,abd,basdon,dirk,new,entropy}, since the phenomenon is interesting both from the theoretical as well as phenomenological points of view. Theoretically, nontrivial issues arise when the emission rate is computed using perturbation theory to the lowest perturbative order: a straightforward computation leads to a wrong result. This was analyzed in detail in~\cite{abd,basdon,dirk,new}, where it was shown that in order to arrive at the correct result, higher order contributions of the electromagnetic interaction must be properly taken into account. A summary of this analysis can be found in \cite{springeremiss}. From the phenomenological point of view, noise-induced radiation emission sets the strongest bound on the gravity-related collapse model~\cite{NaturePhys} and on the collapse parameter $\lambda$ of the CSL model in the region $r_C \leq 10^{-6}$ m~\cite{entropy,ad1,sci,cantilever,cold,diamonds,gravwave} (more details on $\lambda$ and $r_C$ are given below, after Eq. (\ref{Nandg})).

In~\cite{NaturePhys} an accurate analysis of the radiation emission induced by gravity-related collapse was performed, which ruled out the parameter-free version of the model. The two main strengths of that analysis are:
\begin{enumerate}
\item From the theoretical point of view, the emission from the atomic nucleus scales quadratically with the atomic number;
\item From the experimental point of view, to have performed a dedicated experiment with a good knowledge of the setup and of the low background environment.
\end{enumerate}
The analysis done in this paper shares the same strengths for the CSL model. 

In section 2 we introduce the CSL model and discuss the main theoretical result, given by Eq. (\ref{eq:ratefinalevero}), which is derived in the Appendix. In section 3 the data analysis is presented in details and the novel bound on the CSL parameters reported in Fig. \ref{limit}. Finally, in the conclusion, we summarise the main achievement of the paper and suggest possible improvements for future activity in this field.

\section{Theoretical Analysis}

We compute the radiation emission rate for an atomic system according to the CSL model. In the previous literature, this emission rate was obtained by applying the formula derived for the emission rate from free electrons to the outer electrons of the atoms. Precisely, for Germanium atoms, it was first assumed that the relevant contribution comes from the 4 outer electrons of the atom~\cite{fu}, while later the emission from the 30 outer electrons was considered~\cite{entropy}. 

Here we derive a formula for the emission rate that is valid for generic systems, not just for free particles. This derivation shows that there is a relevant regime where also the emission from the nucleons plays an important role. We show that when the energy of the emitted radiation is in the range $10$-$10^5$ keV, all protons in the nucleus emit coherently as the whole nucleus was a single free particle with charge $Q=N_A e$, where $N_A$ is the atomic number of the atom and $e$ the electric charge. On the other hand, the emission from the electrons is incoherent i.e. they emit independently. Since the emission rate is proportional to the square of the charge $Q^2$, this means that the emission from a nucleus grows quadratically with the atomic number $N_A$, while that from the electrons grows linearly with $N_A$. In the case of Germanium, $N_A=32$ which means that the emission is amplified by a factor $32^2 + 32 =1056$, to be compared with the largest amplification factor of 30 considered in the previous literature~\cite{entropy}.

We first introduce the basic formulae which are relevant for the subsequent analysis. The evolution of the state vector $|\phi_{t}\rangle$ in the CSL model is given by~\cite{csl}:
\begin{eqnarray}\label{CSLnonlinear}
d|\phi_{t}\rangle&=&\Big[-\frac{i}{\hbar}\hat{H}dt+\frac{\sqrt{\lambda}}{m_0}\int d\boldsymbol{x}\,\left(\hat{M}(\boldsymbol{x})-\left\langle \hat{M}(\boldsymbol{x})\right\rangle_t \right)dW_{t}(\boldsymbol{x}) \\
&-&\frac{\lambda}{2 m_0^2}\int d\boldsymbol{x}\int d\boldsymbol{x'}\,e^{-\frac{(\boldsymbol{x}-\boldsymbol{x}')^{2}}{4r_{C}^{2}}}\left(\hat{M}(\boldsymbol{x})-\left\langle \hat{M}(\boldsymbol{x})\right\rangle_t \right)\times\nonumber\left(\hat{M}(\boldsymbol{x'})-\left\langle \hat{M}(\boldsymbol{x'})\right\rangle_t \right)dt\Big]|\phi_{t}\rangle,
\end{eqnarray}
where
\begin{equation}\label{Nandg}
\hat{M}\left(\boldsymbol{x}\right):=\underset{j}{\sum} m_j\hat{\psi}^{\dagger}_j\left(\boldsymbol{x}\right)\hat{\psi}_j\left(\boldsymbol{x}\right)
\end{equation}
and $\left\langle \hat{M}(\boldsymbol{x})\right\rangle_t=\langle\phi_t| \hat{M}(\boldsymbol{x})|\phi_t\rangle$.
Here $\hat{\psi}^{\dagger}_j\left(\boldsymbol{x}\right)$ and $\hat{\psi}_j\left(\boldsymbol{x}\right)$ are, respectively, the creation and annihilation operators of a particle of type ``$j$'' at the point $\boldsymbol{x}$. We are considering the mass proportional version of the CSL model, with $m_j$ the mass of a particle of type ``$j$'' and $m_0$ a reference mass, taken equal to the mass of the nucleon \footnote{Here we do not consider the spin degrees of freedom since they give a negligible contribution to the emission rate.}. The noise field $w_t(\boldsymbol{x}):=dW_t(\boldsymbol{x})/dt$, has zero average and correlation
\begin{equation}\label{noiscor}
\mathbb{E}[w(\boldsymbol{x},t)w(\boldsymbol{x}',t')]=e^{-\frac{(\boldsymbol{x}-\boldsymbol{x}')^{2}}{4r_{C}^{2}}}\delta(t-t'),
\end{equation}
with $\mathbb{E}[...]$ denoting the stochastic average. 

The CSL model depends on two parameters: $\lambda$ which sets the strength of the noise and $r_C$, which defines the spatial resolution of the collapse.  
Two values were proposed in the literature for $\lambda$: one, suggested by Ghirardi, Rimini and Weber~\cite{grw}, sets $\lambda_{\textrm{\tiny GRW}}=10^{-16}$ s$^{-1}$ while the other one, proposed by Adler~\cite{ad1}, sets $\lambda_{\textrm{\tiny ADLER}}=10^{-8\pm 1}$ s$^{-1}$. The value $\lambda_{\textrm{GRW}}$ is somehow the smallest possible choice for $\lambda$: for smaller values the effect of the noise would be too small, making the collapse too weak to guarantee an efficient suppression of macroscopic superpositions \cite{toros}. The value $\lambda_{\textrm{\tiny ADLER}}$ is based on the requirement that the process of latent image formation in emulsions yields definite images by collapsing superpositions fast enough. Regarding the correlation length $r_C$, a reference value typically considered in the literature is $r_C=10^{-7}$ m, while previous bounds together with the requirements of having an efficient suppression of macroscopic superpositions require $r_C\geq 10^{-9}$ m \cite{cardon} (see also Fig. \ref{limit}).

The stochastic terms in Eq. (\ref{CSLnonlinear}) imply an acceleration on the particles, which induces the emission of radiation. The computation of the radiation emission rate for a system of $N$ particles with mass $m_j$ and charge $q_j$ is given in the appendix and the final result is:
\begin{equation}\label{ratemigen}
\frac{d\Gamma}{dE}=A\frac{\hbar\lambda}{4\pi^{2}\varepsilon_{0}m_{0}^{2}r_{C}^{2}c^{3}E}\, ,
\end{equation}
where $\hbar$ and $c$ have the usual meaning, $\varepsilon_0$ is the vacuum permittivity and $E$ the energy of the emitted radiation. The amplification factor $A$ takes different forms depending on the kind of emission: in the case of coherent emission one has $A=\left(\sum_{j=1}^N q_{j}\right)^2$, while for incoherent emission $A=\sum_{j=1}^N q_{j}^{2}$ (see Eqs. (\ref{rateinc2}) and (\ref{ratecohreal}) of the appendix).

Whether the radiation emission is coherent or incoerent depends on the relation between the typical distance $L$ among the particles which are emitting radiation, the wavelength $\lambda_k$ of the emitted radiation and the noise correlation length $r_C$. As shown and discussed in detail in the appendix, when the distance between two particles is larger than $\lambda_k$ the emission is always incoherent. On the other hand, particles at a distance smaller than $\lambda_k$ and $r_C$ always emit coherently.

We now apply the result in Eq. (\ref{ratemigen}) to compute the emission rate from an atom. 
To maximize the emission rate, one needs to find a region where the wavelength of the emitted radiation $\lambda_k$ is small enough to guarantee that the electrons and the nucleus emit independently from each other, so that their contribution does not add coherently to zero, since they have opposite charges. Therefore one has to consider wavelengths $\lambda_k\ll 0.1$ nm, which is the typical distance between electrons and the nucleus. 
On the other hand, it is convenient to take the opposite limit when considering the emission from the nucleus. In the nucleus, all protons have the same positive charge, therefore the biggest effect is obtained in the regime where they emit coherently. This requires $\lambda_k, r_C\gg 10^{-6}$ nm, which is the typical size of nucleus.  

Since values of $r_C$ smaller than 1 nm are already excluded by previous bounds \cite{cardon}, these conditions are fulfilled by taking radiation with wavelengths $\lambda_k\sim 10^{-6}\,-\,10^{-1}$ nm, which corresponds to energies in the range $E\sim 10\,-\,10^{5}$ keV. 
In this case, the emission rate from multiple atoms is computed according to the following formula:
\begin{equation}\label{eq:ratefinalevero}
\frac{d\Gamma}{dE}=N_{atoms}\times \left(N_{A}^{2}+N_{A}\right)\times\frac{\lambda \hbar e^{2}}{4\pi^{2}\varepsilon_{0}m_{0}^{2}r_{C}^{2}c^{3}E}
\end{equation}
with $N_{A}$ the atomic number. The factor $N_{A}^2$ is due to the coherent emission from the nuclei, while the linear contribution $N_{A}$ comes from the electrons. Note that the linear contribution due to the electrons can be taken into account only when the energy of the emitted photons is in the range $E\sim 10\,-\,100$ keV. In fact, for larger energies, the electrons are in the relativistic regime where our results do not apply.

\section{Experimental Results}
\begin{figure}
\centering
\includegraphics[width=0.5\textwidth]{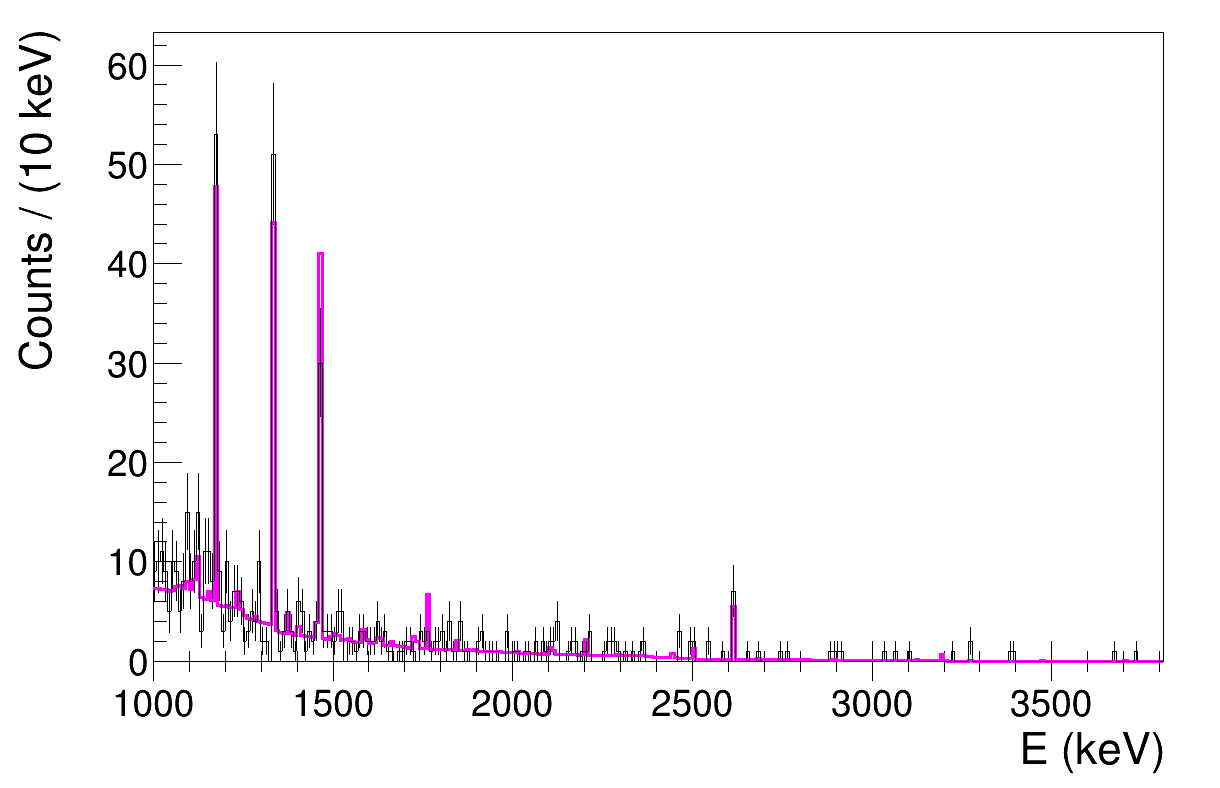}
\caption{Measured X-rays spectrum (black histogram) and simulated background distribution (magenta histogram) in the energy range $\Delta E$ = (1000$\div$3800) keV.}
\label{data_mc}
\end{figure}

A measurement was performed to constraint the parameters of the CSL model, by comparing the X-ray spectrum acquired with an ultra-pure Germanium detector with the expected noise-induced radiation emitted by the materials of the experimental apparatus. 
The setup, based on a High Purity Germanium (HPGe) coaxial p-type detector surrounded by layers of electrolytic copper and pure lead shielding, is extensively described in Ref. \cite{NaturePhys}. Data was acquired in the extremely low background environment of the Gran Sasso underground National Laboratory of INFN (LNGS-INFN), which is characterised by an average overburden of 3800 meters water equivalent, see Ref. \cite{lngs}. 
The measured spectrum, corresponding to an exposure of 124 kg $\cdot$ day, is shown in black in Fig. \ref{data_mc}.

The aim of the data analysis for this measurement is to obtain the probability distribution function (\emph{pdf}) of the expected number of measured counts (derived in Section \ref{ul}), which is a function of the parameters $\lambda$ and $r_C$ of the model, from which an unambiguous limit can be extracted. To this end the expected contribution of the spontaneous radiation signal is evaluated by weighting the theoretical rate (Eq. (\ref{eq:ratefinalevero})) with the detection efficiencies, as described in Section \ref{sig}; the contribution of the background, which is mainly due to the emission of residual radionuclides present in the setup constituents, is also estimated (see Section \ref{bkg}).    
In order to achieve a complete characterisation of the experimental setup, to calculate the detection efficiencies and to simulate the background contribution, the whole detector was described by a validated Monte Carlo (MC) code (Ref. \cite{boswell2010textsc}) based on the GEANT4 software library (Ref. \cite{agostinelli2003geant4}).

\subsection{Background simulation}\label{bkg}

The MC simulation of the background takes as input the measured activities of each radionuclide 
detected in each component of the setup; the simulation accounts for the emission probabilities and the decay schemes. The photon propagation and interactions inside the materials of the detector, which give rise to the continuum part of the background spectrum, are realistically described. Finally, the detection efficiencies are considered.

The measured and simulated spectra are in good agreement in the range
($\Delta E = (1000 \div 3800) \mbox{keV}$) as shown in Fig. \ref{data_mc}, where the background is represented by the magenta distribution. The simulation accounts for about 88$\%$ of the measured spectrum. The total number of simulated background counts in $\Delta E$ is $z_b=506$, to be compared with a total number of measured photons $z_c=576$.
All the assumptions of the calculation leading to the theoretical spontaneous emission rate in Eq. (\ref{eq:ratefinalevero}) are satisfied in  $\Delta E$, provided that only the dominant contribution of protons is considered, since electrons are relativistic in this range.

\subsection{Calculation of the expected signal shape}\label{sig}

\begin{figure*}
    \centering
    \includegraphics[width=0.33\textwidth]{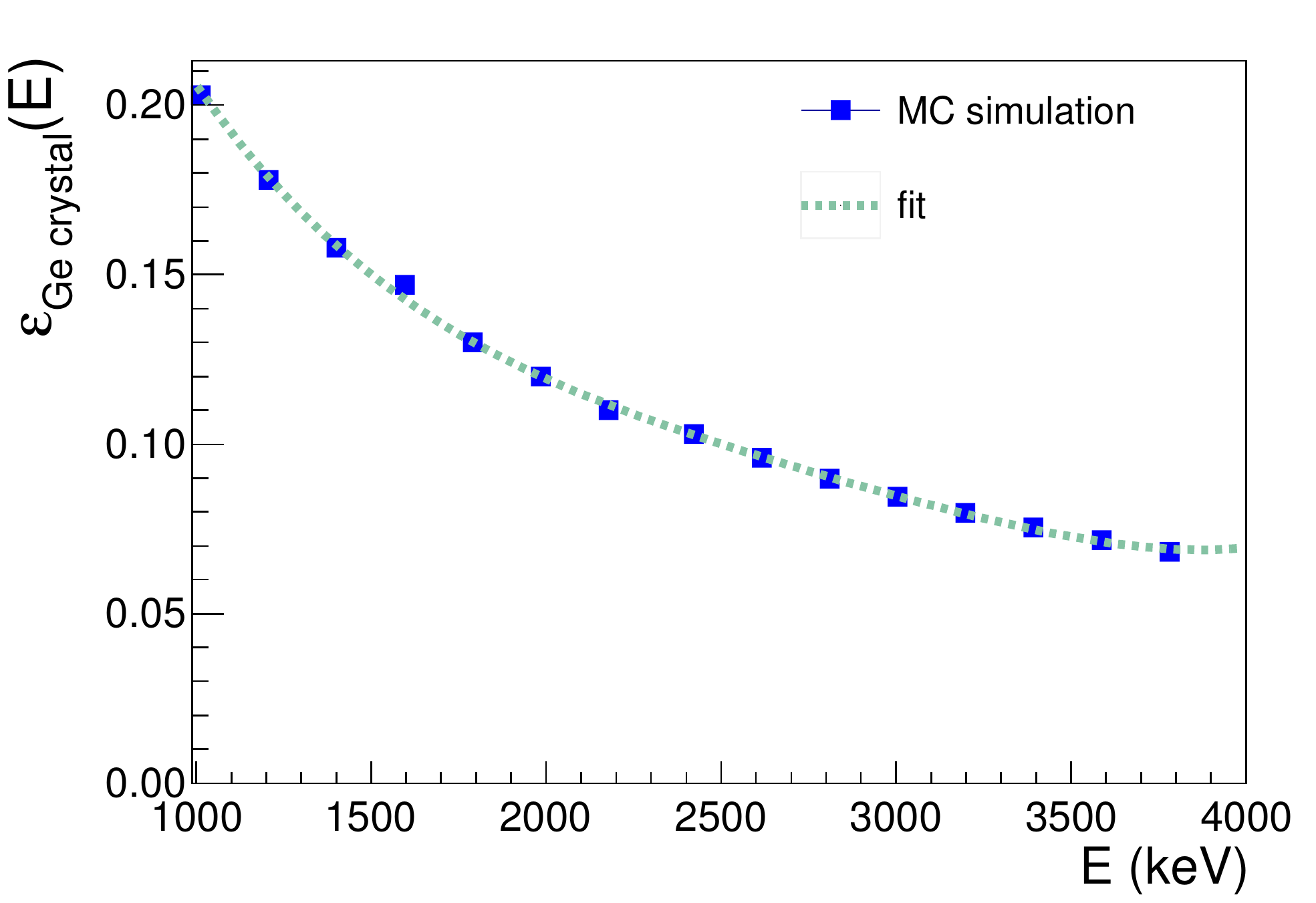}
    \includegraphics[width=0.33\textwidth]{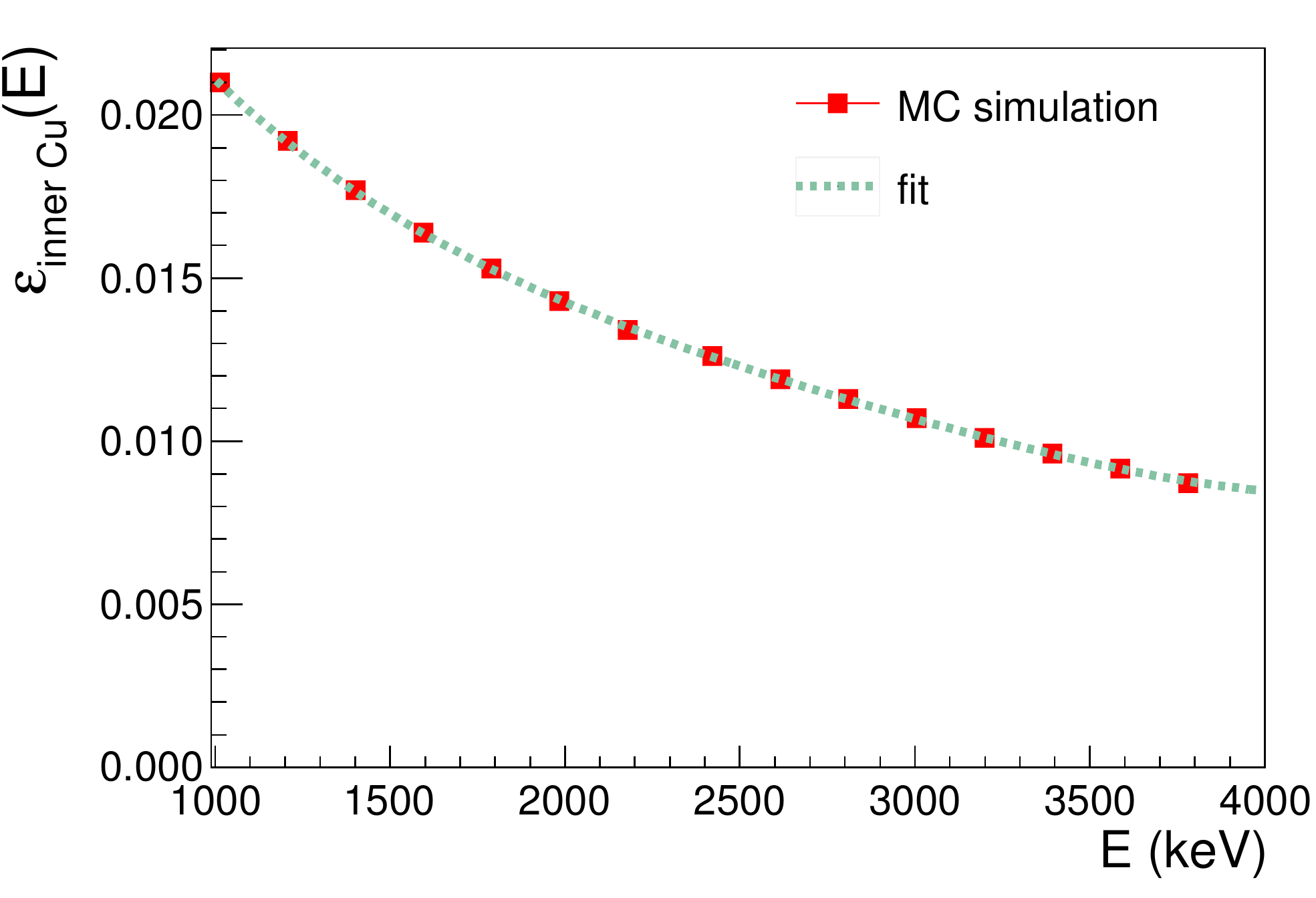}
    \includegraphics[width=0.33\textwidth]{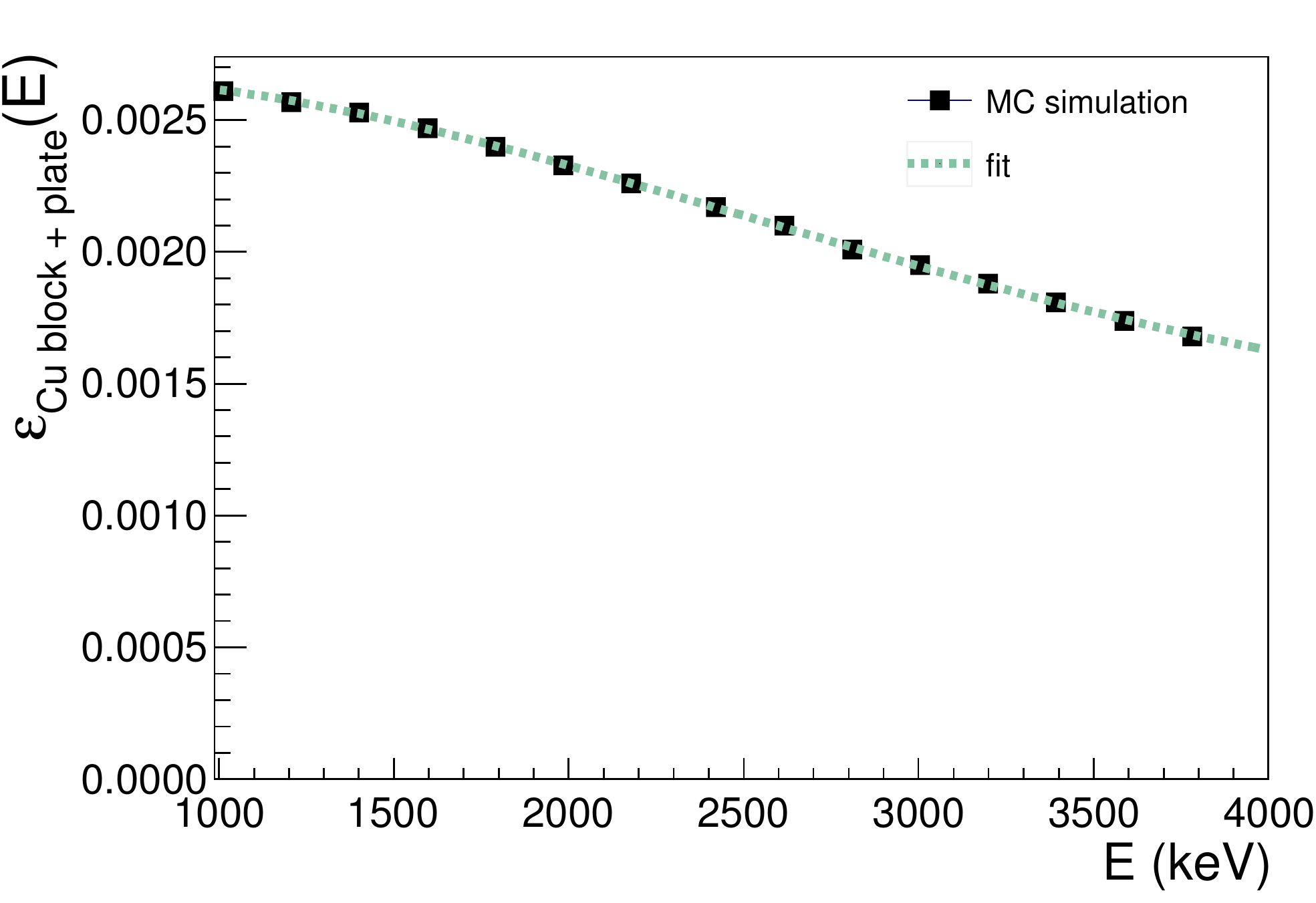}
 \includegraphics[width=0.33\textwidth]{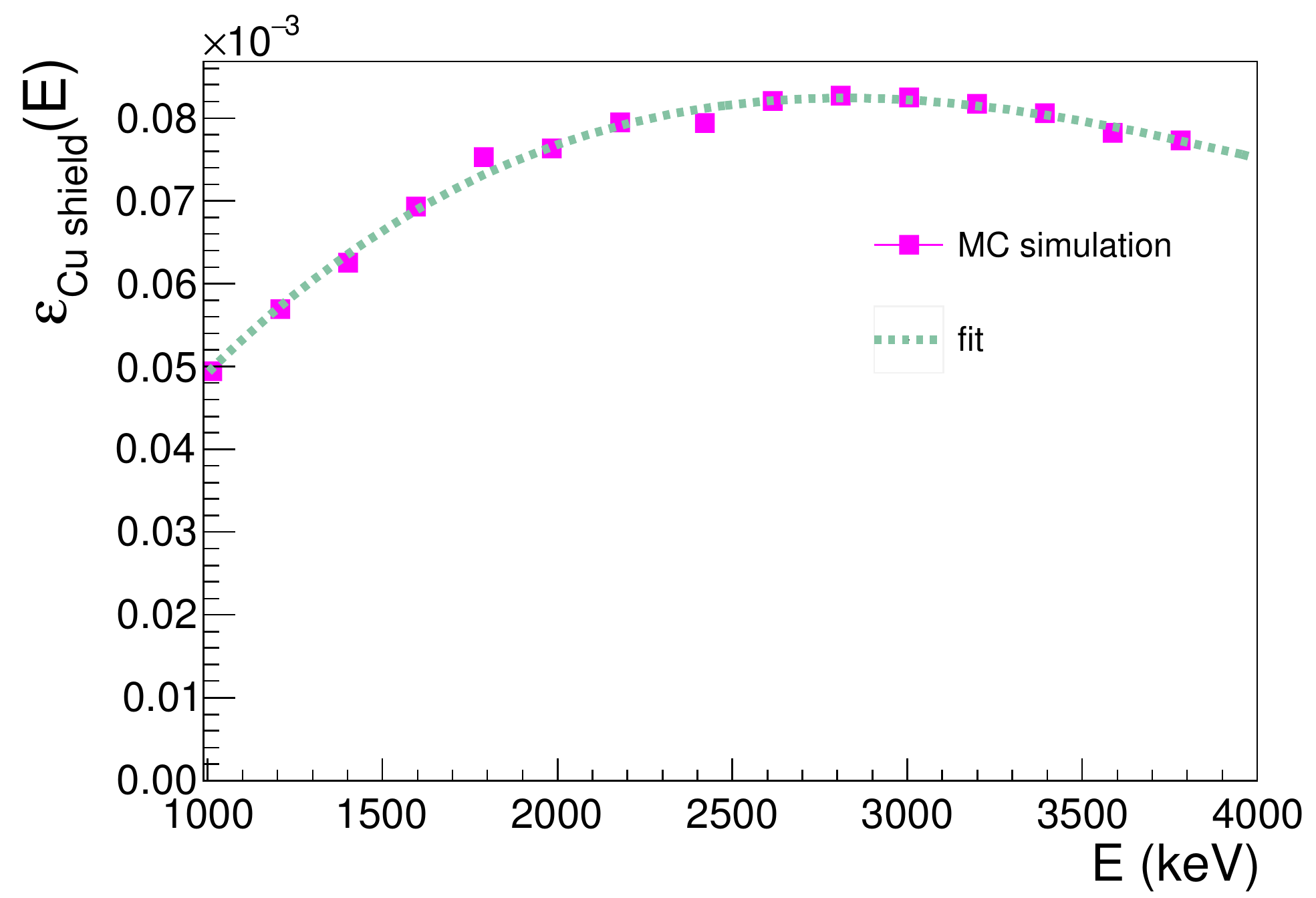}
    \includegraphics[width=0.33\textwidth]{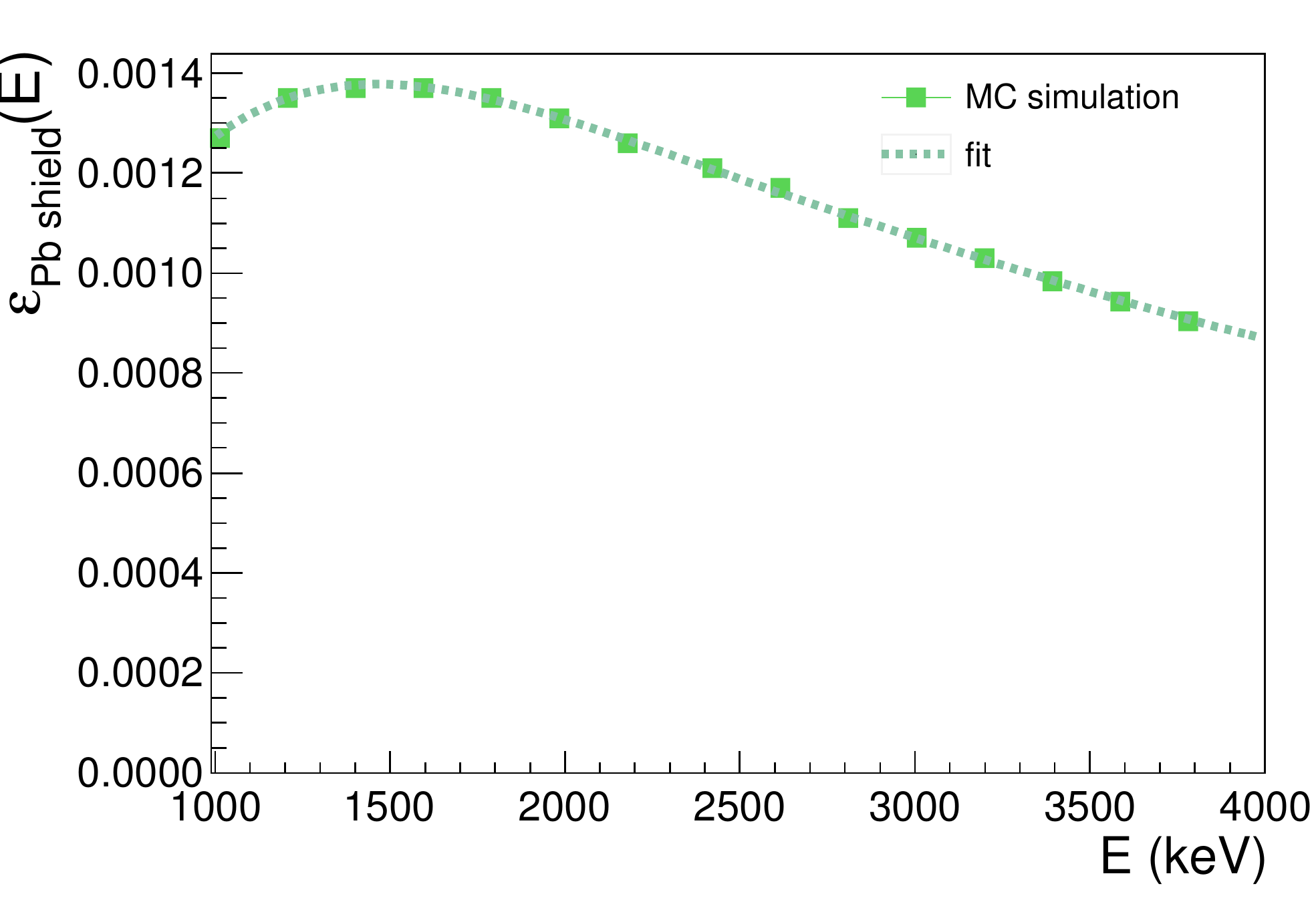}
\caption{\em The results of the fits of the efficiency spectra, for those components of the setup which are found to give significant contribution: Ge crystal (top left panel), inner Cu (top middle panel), Cu block + plate (top right panel), Cu shield (bottom left panel), Pb shield (bottom right panel). A detailed description of the setup is given in Fig. 2 of Ref \cite{NaturePhys}.}
\label{efficiencies}
\end{figure*}

In order to evaluate the expected spontaneous radiation signal contribution (i.e. the number of spontaneously emitted photons, generated by each atom of each material of the detector, which would be detected by the HPGe in the acquisition time $T$) the theoretical rate is weighted with the detection efficiency function.
Since, due to self-absorption, the efficiency function strongly depends on the geometry of the apparatus, the efficiency is obtained by means of a MC simulation. A scan is performed in steps of 200 eV in the range $\Delta E$, by generating, with a uniform spatial distribution, 10$^8$ photons for each energy and for each component of the apparatus (we refer to Fig. 2 of Ref \cite{NaturePhys} for a detailed description of the setup). 
The efficiency spectra, for all the materials contributing significantly, are shown in Fig. \ref{efficiencies}.
The efficiency function (for the $i$-th material of the apparatus) is then the result of a polynomial fit to the distribution resulting from the simulation:
\begin{equation}\label{sup2}
\epsilon_i(E)=\sum_{j=0}^{c_i} \xi_{ij}E^j,
\end{equation}
$c_i$ is the degree of the polynomial expansion, $\xi_{ij}$ is the matrix of the coefficients.
The results of the fits are also shown in Fig. \ref{efficiencies}, and the corresponding parameters are summarised in Table \ref{tab:fit_parameters}, with the corresponding statistical errors.

\begin{table*}
    \centering
    \resizebox{1\columnwidth}{!}{%
    \begin{tabular}{c|c|c|c|c|c}
             $i=$ & $\mathrm{Ge\ crystal}$ & $\mathrm{Inner\ Cu}$ & $\mathrm{Cu\ block\ +\ plate}$ & $\mathrm{Cu\ shield}$ & $\mathrm{Pb\ shield}$\\
              \hline\\
              \hline
        $\xi_{i0}$ &  (4.82 $\pm$ 0.03) $\cdot 10^{-1}$  &  (3.77 $\pm$ 0.04) $\cdot 10^{-2}$ & (2.6 $\pm$ 0.1) $\cdot 10^{-3}$  & (-1.01 $\pm$ 0.07) $\cdot 10^{-5}$ & (-5.76 $\pm$ 0.03) $\cdot 10^{-4}$  \\ 
        $\xi_{i1}$ &  (-4.42 $\pm$ 0.03) $\cdot 10^{-4}$ & (-2.48 $\pm$ 0.03) $\cdot 10^{-5}$  & (2.9 $\pm$ 1.4) $\cdot 10^{-7}$ &  (7.8 $\pm$ 0.1) $\cdot 10^{-8}$ & (3.812 $\pm$ 0.003) $\cdot 10^{-6}$ \\ 
        $\xi_{i2}$ &  (2.10 $\pm$ 0.01) $\cdot 10^{-7}$ & (1.03 $\pm$ 0.01) $\cdot 10^{-8}$  &  (-3.1 $\pm$ 0.5) $\cdot 10^{-10}$ &  (-2.07 $\pm$ 0.06) $\cdot 10^{-11}$ &  (-2.728 $\pm$ 0.001) $\cdot 10^{-9}$ \\
        $\xi_{i3}$ &  (-4.87 $\pm$ 0.03) $\cdot 10^{-11}$ &  (-2.24 $\pm$ 0.04) $\cdot 10^{-12}$ & (5.7 $\pm$ 1.6) $\cdot 10^{-14}$ & (1.61 $\pm$ 0.09) $\cdot 10^{-15}$ &  (9.036 $\pm$ 0.004) $\cdot 10^{-13}$ \\
        $\xi_{i4}$ &  (4.32 $\pm$ 0.07) $\cdot 10^{-15}$  & (1.93 $\pm$ 0.08) $\cdot 10^{-16}$  & (-3.1 $\pm$ 3.3) $\cdot 10^{-18}$  &  - &  (-1.477 $\pm$ 0.001) $\cdot 10^{-16}$ \\
        $\xi_{i5}$ &  -  & -  & -  & -  & (9.60 $\pm$ 0.02) $\cdot 10^{-21}$ \\
    \end{tabular}
    }
    \caption{\em The table summarises the parameters obtained from the best fit to the efficiency spectra, for each component of the setup which gives a significant contribution.
}
    \label{tab:fit_parameters}
\end{table*}

The expected signal contribution is then given by:
\begin{equation}\begin{split}\label{sup3}
z_s\left( \frac{\lambda}{r_C^2} \right) & = \int_{\Delta E} \sum_i \left. \frac{d\Gamma}{dE} \right|_i \epsilon_i (E) \, dE = \int_{\Delta E} \sum_i N^2_{pi} \, \alpha_i \, \beta \, \frac{\lambda}{r_C^2 E} \, \sum_{j=0}^{c_i} \xi_{ij}E^j \, dE  =\\
&= 2.0986 \ \frac{\lambda}{r_C^2} = a \frac{\lambda}{r_C^2}
\end{split}\end{equation}
where the sum extends over all the components of the setup, $N_{pi}$ is the number of protons in the atoms of the $i$-th material. $\alpha_i$ is defined as:
\begin{equation}\label{sup4}
\alpha_i = m_i \, n_i \, T,
\end{equation}
with $n_i$ the number of atoms per mass unit of the $i$-th material. According to Eq. (\ref{eq:ratefinalevero}) $\beta$ is defined as:
\begin{equation}\label{sup5}
\beta = \frac{\hbar e^2}{4\pi^2 \epsilon_0 c^3 m_0^2}.
\end{equation}

In Eq. \eqref{sup3} the explicit dependence of the expected signal contribution on the parameters of the model appears. The expected signal shape as a function of the energy, namely the integrand in Eq. \eqref{sup3}, is shown in Fig. \ref{plotsig}.

\subsection{Novel limit on the CSL parameters}\label{ul}

The stochastic variable $z_c$ is distributed according to a Poissonian:

\begin{equation}\label{sup6}
p(z_{c}|\Lambda_c)=\frac{\Lambda_{c}^{z_{c}}e^{-\Lambda_{c}}}{z_{c}!};
\end{equation}
$\Lambda_c$ is the expected value for the total number of measured photons in $\Delta E$. 

Since the total number of signal and background counts are also Poissonian variables, $\Lambda_c$ can be rewritten as the sum of the background and of the signal contribution to spontaneously emitted photons:

\begin{equation}\label{sup7}
\Lambda_c \left( \frac{\lambda}{r_C^2} \right) = \Lambda_b+\Lambda_s \left( \frac{\lambda}{r_C^2} \right),
\end{equation}
with $\Lambda_s=z_s+1$ and $\Lambda_b=z_b+1$. The $pdf$ of $\Lambda_c\left( \frac{\lambda}{r_C^2} \right) $ is then obtained from Eq. \eqref{sup6} by applying the Bayes theorem:

\begin{figure}
\centering
\includegraphics[width=0.4\textwidth]{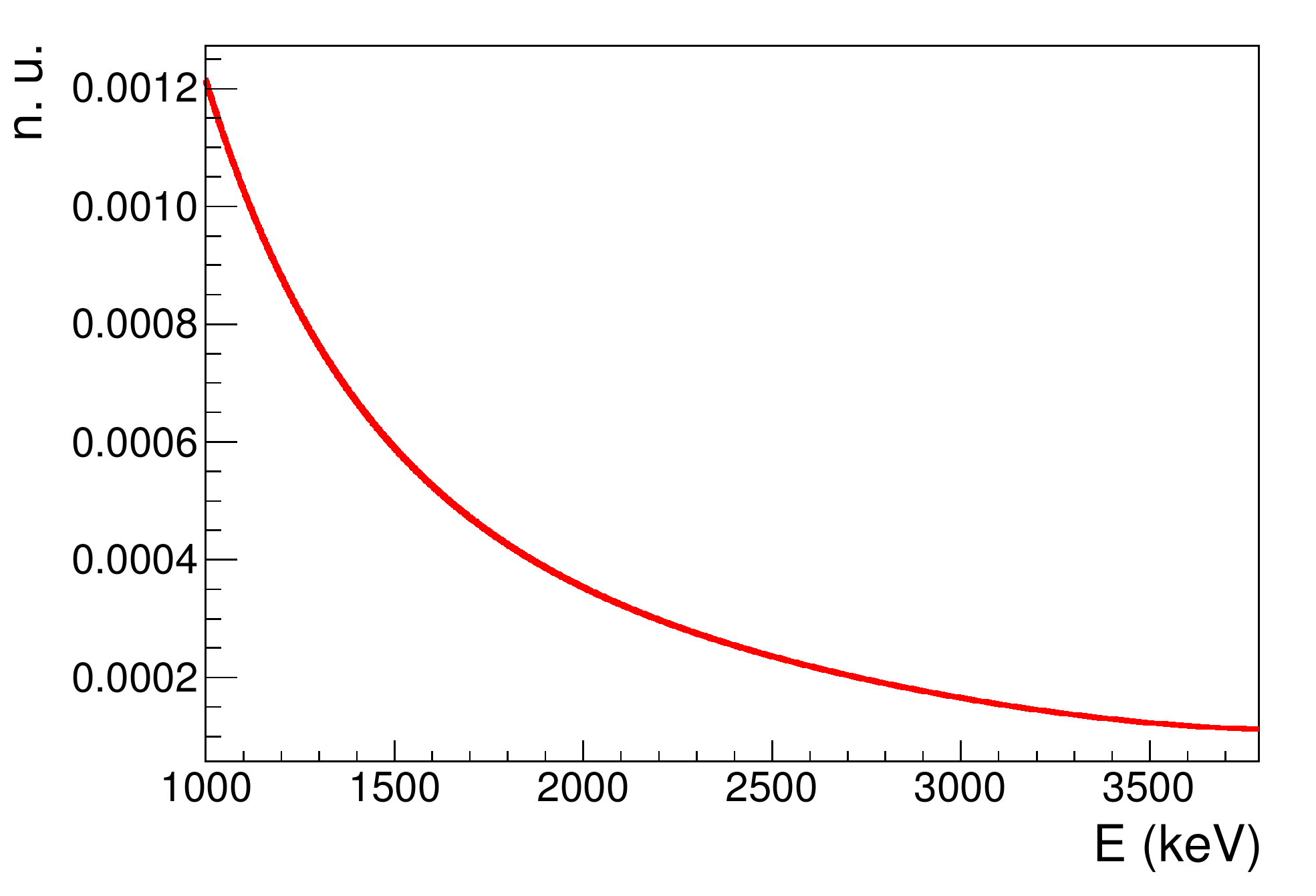}

\caption{{\it 
Energy distribution of the expected spontaneous radiation signal, resulting from the sum of the emission rates of all the components of the setup, weighted with the efficiency functions. The area of the distribution is normalised to unity (n.u.) in the energy range $\Delta E$ = (1000$\div$3800) keV.} }
\label{plotsig}
\end{figure}
\begin{equation}\label{sup8}
\tilde{p}\left(\Lambda_c \left(\left. \frac{\lambda}{r_C^2} \right)  \right| p(z_{c}|\Lambda_c)\right) = \frac{p(z_{c}|\Lambda_c) \cdot \tilde{p}_0(\Lambda_c\left( \frac{\lambda}{r_C^2} \right) )}{\int_D p(z_{c}|\Lambda_c) \cdot \tilde{p}_0(\Lambda_c\left( \frac{\lambda}{r_C^2} \right) ) \, d\Lambda_c}.
\end{equation} 
If a uniform prior is assumed the posterior distribution function is:
\begin{equation}\label{sup9}
\tilde{p}\left(\Lambda_c\left(\left. \frac{\lambda}{r_C^2} \right)\right|p(z_{c}|\Lambda_c)\right) = \frac{[\Lambda_{c}\left( \frac{\lambda}{r_C^2} \right)]^{z_{c}} \, e^{-\Lambda_{c}}}{\Gamma(z_c+1)}.
\end{equation} 
We are now in the position to set an upper bound on the $\lambda$ parameter of the model. Given the linear relation among $\Lambda_c$ and $\lambda$, this can be easily derived by using the cumulative $pdf$ of $\Lambda_c$. In particular, by setting $r_C = 10^{-7}$ m (as originally proposed in Ref. \cite{grw}), and solving the integral equation:
\begin{equation}\label{sup10}
\tilde{P}\left( \bar{\Lambda}_c \right) = \frac{\int_0^{\bar{\Lambda}_c}[\Lambda_{c}\left( \frac{\lambda}{r_C^2} \right)]^{z_{c}} \, e^{-\Lambda_{c}} d\Lambda_c}{\Gamma(z_c+1)}=\frac{\gamma(z_c+1,\bar{\Lambda}_c)}{\Gamma(z_c+1)}=0.95.
\end{equation}
We obtain the upper limit on $\lambda$ 
\begin{equation}
\lambda < 5.2 \cdot 10^{-13}\; \textrm{s}^{-1}
\end{equation}
with a probability of 0.95. 
This improves the previous best result available in literature (Eq. (18) in Ref. \cite{entropy}) by a factor 13. 

\begin{figure}
\centering
\includegraphics[width=0.45\textwidth]{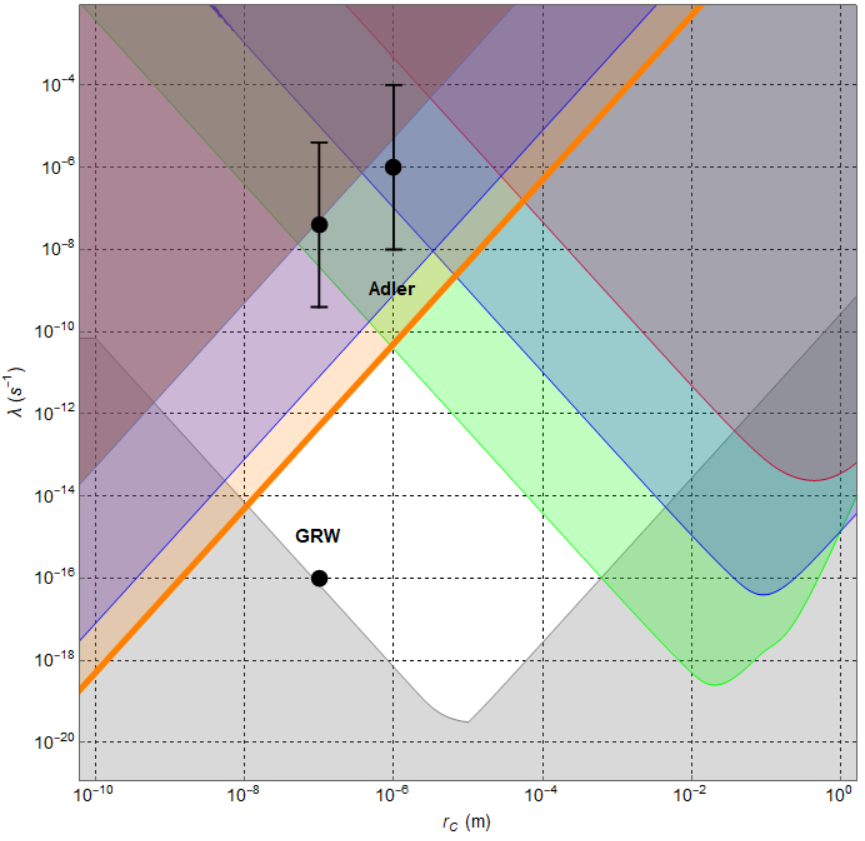}
\caption{Mapping of the $\lambda - r_C$ CSL parameters: the proposed theoretical values (GRW \cite{grw}, Adler \cite{ad1}) are shown as black points and the region excluded by theory is represented in gray \cite{toros}. The right part of the parameter space is excluded by the bounds coming from the study of gravitational waves detectors: Auriga (red), Ligo (Blue) and Lisa-Pathfinder (Green) \cite{gravwave}. On the left part of the parameter space there is the bound from the study of the expansion of a Bose-Einstein condensate (red) \cite{cold} and the most recent from the study of radiation emission from Germanium (purple) \cite{entropy}. This bound is improved by a factor 13 by this analysis performed here, with a confidence level of 0.95, and it is shown in orange.
}
\label{limit}
\end{figure}

More in general an exclusion plot, in the $\lambda - r_C$ plane, can be obtained from Eq. \eqref{sup10}
by varying $r_C$. Fig. \ref{limit} shows the $\lambda - r_C$ mapping, where the region which is excluded by this analysis, with a probability of 0.95, is represented in orange.  
We obtained the most stringent limits on the CSL collapse parameters in a broad range of the parameters space (see also Ref. \cite{cardon} for comparison with other experimental constraints).

\section{Conclusions and perspectives}
We set the strongest bounds on CSL parameters in the region $r_C\leq 10^{-6}$ m by refining, on one hand,  the theoretical calculations, considering the realistic emission of spontaneous radiation from atoms and in particular the coherent emission from nuclei, which leads to a quadratic amplification of the radiation emission rate with the atomic number of the emitting atoms. On the other hand, we performed a dedicated experiment searching for this radiation with a High Purity Germanium detector at the LNGS-INFN Gran Sasso underground laboratory, which was complemented by a refined work on the simulation of the background coming from known sources and by a statistical analysis. All these improvements allowed us to gain a factor 13 in the bounds set on the CSL model parameters with respect to previous similar searches. This result also establishes an absolute record on $\lambda$-rc mapping of the CSL parameters below $r_C=10^{-6}$ m, proving the method of searching for the spontaneous radiation predicted by CSL models as being, today, an unsuprassable one.

Baring on the success of this combined theoretical and experimental approach, for the future we plan, on one side, to develop  
new methods for testing non-Markovian and dissipative collapse models, and, on the other, to refine our experimental techniques as well as analyses methods by implementing Machine Learning algorithms. In such a way we shall extend our search to a broader energy range where we are able to  appreciate features of realistic CSL models towards a deeper understanding of their dynamics and, more in general, of quantum physics and its possible scientific and technological applications limitations

\section*{\small Acknowledgments}

S. D. acknowledges support from INFN. 
A. B. acknowledges financial support from the H2020 FET Project TEQ (Grant No. 766900), the Foundational Questions Institute and Fetzer Franklin Fund, a donor advised fund of Silicon Valley Community Foundation (Grant No. FQXi-RFP-CPW- 2002), INFN and the University of Trieste.
C. C. acknowledges support from the H2020 FET TEQ (Grant No.  766900), INFN (VIP) and the Foundational Questions Institute (ICON project). K. P. acknowledges support from the Centro Ricerche Enrico Fermi - Museo Storico della Fisica e Centro Studi e Ricerche ``Enrico Fermi'' (Open Problems in Quantum Mechanics project). \\
The authors wish express their deep gratitude to S.L. Adler for many enjoyable and very stimulating discussions on this topic. The authors thanks M. Carlesso for sharing the Mathematica code used to generate Fig. 4.

\section*{\Large Appendix: Theoretical analysis}

Even though our goal is to compute the radiation emission rate from a Germanium crystal, we perform a more general theoretical analysis considering a generic system made of charged $N$ particles and we ask for the probability of a photon being  emitted due to the interaction of the particles with the noise. 
For such a system, the CSL master equation can be obtained from Eq. (\ref{CSLnonlinear}) projecting to the $N$-particles Hilbert spaces and it reads: 
\begin{equation}\label{ME}
\frac{d\hat{\rho}_{t}}{dt}=-\frac{i}{\hbar}\left[\hat{H},\hat{\rho}_{t}\right]-\frac{\lambda}{2m_{0}^{2}}\!\!\int \!\!d\boldsymbol{x}\!\!\int d\boldsymbol{x}'\!\!e^{-\frac{(\boldsymbol{x}-\boldsymbol{x}')^{2}}{4r_{C}^{2}}}\left[\hat{\mu}(\boldsymbol{x}),\left[\hat{\mu}(\boldsymbol{x}'),\hat{\rho}_{t}\right]\right]\nonumber
\end{equation}
where $\hat{\rho}_t=\mathbb{E}[|\phi_{t}\rangle\langle \phi_{t}|]$, $\hat{\mu}(\boldsymbol{x})=\hat{\mu}(\boldsymbol{x},\hat{\boldsymbol{x}}_{1},...,\hat{\boldsymbol{x}}_{N})=\sum_{j}m_{j}\delta(\hat{\boldsymbol{x}}_{j}-\boldsymbol{x})$ with $\hat{\boldsymbol{x}}_{j}$ the position operator of the $j$-th particle.

It is straightforward to prove that the master equation \eqref{ME} can be unravelled by the following stochastic Schr\"odinger equation:
\begin{equation}
i\hbar\frac{d|\psi_{t}\rangle}{dt}=\left[\hat{H}+V(\hat{\bm{x}}_{1},...,\hat{\boldsymbol{x}}_{N}, t)\right]|\psi_{t}\rangle,\label{eq:unrav}
\end{equation}
where
\begin{equation}\label{potential_new}
V(\hat{\boldsymbol{x}}_{1},...,\hat{\boldsymbol{x}}_{N}, t)=-\frac{\hbar\sqrt{\lambda}}{m_{0}}\int d\boldsymbol{x}\hat{\mu}(\boldsymbol{x},\hat{\boldsymbol{x}}_{1},...,\hat{\boldsymbol{x}}_{N})w(\boldsymbol{x},t).
\end{equation}
Eq. (\ref{eq:unrav}) does \textit{not} describe any collapse
of the wave function, being a standard
linear Schr\"odinger equation.  
Yet, as long as we are interested in computing averaged quantities, it is equally good---and mathematically easier to handle---than directly using the master equation (for an example of a calculation of spontaneous photon emission process in the CSL model performed directly from the master equation, see \cite{dirk}). 

We are now in position to derive Eq.~(\ref{ratemigen}) of the main text, which
gives the power emission formula from a generic rigid body. Several studies of the radiation emitted by charged particles according to the CSL model were carried on in the past \cite{fu, ar, abd, dirk} and in \cite{bd,basdon} for the Quantum Mechanics with Universal Position Localization (QMUPL) model \cite{ref:qmupl} but they mainly focused on the emission from free electrons. Here we focus on the emission from all particles in the atom, including the contribution of protons, which will turn out to be dominant. We use a semi-classical approach, where the power emitted is directly linked to the acceleration of the particles of the system. This approach was proved to be valid in previous articles, where the semiclassical calculation of the emission rate from a free particle was performed and the result was perfectly equivalent to that obtained using a fully quantum mechanical calculation \cite{ad1} (see Appendix A of the paper).  

The starting point is the classical formula for the total power emitted by a system of charged particles \cite{griff}:
\begin{equation}
P(t)=R_{sp}^{2}\int d\Omega\,S(R_{sp}\hat{\boldsymbol{n}},t),
\label{eq:powS}
\end{equation}
where $S(R_{sp}\hat{\boldsymbol{n}},t)$ is the magnitude of the Poynting
vector at time $t$, at point $R_{sp}\hat{\boldsymbol{n}}$ on a spherical surface of radius $R_{sp}$ (upon which the
integration is performed) along the direction given by the unit vector $\hat{\boldsymbol{n}}$.
The magnitude of Poynting vector is  proportional to the square of the electric field:  
\begin{equation}
S(\boldsymbol{r},t)=\varepsilon_{0}c \boldsymbol{E}^{2}(\boldsymbol{r},t).
\end{equation}
The total electric field generated by $N$ charged particles is:
\begin{equation}
\boldsymbol{E}(\boldsymbol{r},t)=\sum_{i=1}^{N}\boldsymbol{E}_{i}(\boldsymbol{r},t)
\end{equation}
where, in the non-relativistic limit, the radiative part of the electric field of each particle is \cite{griff}
\begin{equation}
\boldsymbol{E}_{i}(\boldsymbol{r},t)=\frac{q_{i}}{4\pi\varepsilon_{0}}\left\{ \frac{\hat{\boldsymbol{R}}_{i}(t_{i}')}{R_{i}(t_{i}')}\times\frac{1}{c^{2}}\left[\hat{\boldsymbol{R}}_{i}(t_{i}')\times\ddot{\boldsymbol{r}}_{i}(t_{i}')\right]\right\} =\frac{q_{i}}{4\pi\varepsilon_{0}c^{2}R_{i}(t_{i}')}\left\{ \hat{\boldsymbol{R}}_{i}(t_{i}')[\ddot{\boldsymbol{r}}_{i}(t_{i}')\cdot\hat{\boldsymbol{R}}_{i}(t_{i}')]-\ddot{\boldsymbol{r}}_{i}(t_{i}')\right\} 
\label{eq:Ei1}
\end{equation}
where $t_{i}':=t-\frac{|\boldsymbol{r}-\boldsymbol{r}_{i}(t_{i}')|}{c}$ is the retarded time of the $i$-th particle and $\boldsymbol{R}_{i}(t_{i}'):=\boldsymbol{r}-\boldsymbol{r}_{i}(t_{i}')$ with $\boldsymbol{r}_{i}(t_{i}')$ the position of the $i$-th particle at its retarded time $t_{i}'$. 

Consistently with taking only the radiative part of the electric field generated by each particle, we assume that the points where the field is evaluated are very far from the system, i.e. the integration in Eq. (\ref{eq:powS}) is performed over a sphere with radius $R_{sp}\gg\boldsymbol |{\boldsymbol{r}}_{i}(t_{i}')|$.
Then we can approximate each particle position as if they are all at the origin i.e. $R_{i}\simeq r$, $\hat{\boldsymbol{R}}_{i}(t_{i}')\simeq\hat{\boldsymbol{n}}$ and accordingly Eq.~(\ref{eq:Ei1}) can be approximated as
\begin{equation}\label{eq:Ei11}
\boldsymbol{E}_{i}(\boldsymbol{r},t)\simeq\frac{q_{i}}{4\pi\varepsilon_{0}c^{2}r}\left[\hat{\boldsymbol{n}}[\ddot{\boldsymbol{r}}_{i}(t_{i}')\cdot\hat{\boldsymbol{n}}]-\ddot{\boldsymbol{r}}_{i}(t_{i}')\right].
\end{equation}
We wish to remark that, despite the approximations from Eq.~(\ref{eq:Ei1}) to Eq.~(\ref{eq:Ei11}), there is still an implicit account for the different particles positions in the retarded times $t_{i}'$, which will play a fundamental role in the following analysis.
 
Since we will eventually compute the emission rate at a given frequency $\omega$,
it is useful to introduce the Fourier transform of the electric fields
\begin{equation}
\boldsymbol{E}_{i}(\boldsymbol{r},\omega)=\frac{q_{i}}{4\pi\varepsilon_{0}c^{2}r}\int_{-\infty}^{+\infty}dt\,e^{-i\omega t}\left[\hat{\boldsymbol{n}}[\ddot{\boldsymbol{r}}_{i}(t_{i}')\cdot\hat{\boldsymbol{n}}]-\ddot{\boldsymbol{r}}_{i}(t_{i}')\right]\label{eq:Eiw}
\end{equation}
and make the change of variable from $t$ to the retarded time
$=t-\frac{|\boldsymbol{r}-\boldsymbol{r}_{i}(t_{i}')|}{c}\simeq t-\frac{|\boldsymbol{r}-\bar{\boldsymbol{r}}_{i}|}{c}$. In the last step we assumed that the positions
of the particles of the system do not change too much with respect to their relative distances, and so they can be replaced by averaged positions $\bar{\boldsymbol{r}}_{i}$ (note that this assumption is equivalent to assuming that the system is a rigid body). Then Eq. (\ref{eq:Eiw}) becomes
\begin{align}
\boldsymbol{E}_{i}(\boldsymbol{r},\omega)&\!\simeq\!\frac{q_{i}}{4\pi\varepsilon_{0}c^{2}r}\!\!\!\int_{-\infty}^{+\infty}\!\!\!\!dt e^{-i\omega\left(t+\frac{|\boldsymbol{r}-\bar{\boldsymbol{r}}_{i}|}{c}\right)}\left[\hat{\boldsymbol{n}}[\ddot{\boldsymbol{r}}_{i}(t)\cdot\hat{\boldsymbol{n}}]-\ddot{\boldsymbol{r}}_{i}(t)\right]\nonumber\\
&\!=\!\frac{q_{i}}{4\pi\varepsilon_{0}c^{2}r}e^{-i\omega\frac{|\boldsymbol{r}-\bar{\boldsymbol{r}}_{i}|}{c}}\left[\hat{\boldsymbol{n}}[\ddot{\boldsymbol{r}}_{i}(\omega)\cdot\hat{\boldsymbol{n}}]-\ddot{\boldsymbol{r}}_{i}(\omega)\right]
\end{align}
We can further approximate the exponent using the fact that $r\gg\bar{r}_{i}$, which implies $|\boldsymbol{r}-\bar{\boldsymbol{r}}_{i}|\simeq r-\bar{\boldsymbol{r}}_{i}\cdot\hat{\boldsymbol{n}}$, in order to get:
\begin{equation}
\boldsymbol{E}_{i}(\boldsymbol{r},\omega)\simeq\frac{q_{i}}{4\pi\varepsilon_{0}c^{2}r}e^{-i\omega\frac{r-\bar{\boldsymbol{r}}_{i}\cdot\hat{\boldsymbol{n}}}{c}}\left[\hat{\boldsymbol{n}}[\ddot{\boldsymbol{r}}_{i}(\omega)\cdot\hat{\boldsymbol{n}}]-\ddot{\boldsymbol{r}}_{i}(\omega)\right].
\end{equation}

Going back to Eq.~(\ref{eq:powS}) for the  emitted power, one gets:
\begin{align}
P(t)&=\frac{R_{sp}^{2}\varepsilon_{0}c}{(2\pi)^2}\int_{-\infty}^{+\infty}d\omega\,\int_{-\infty}^{+\infty}d\nu\,e^{i(\omega+\nu)t} \int d\Omega \sum_{i,j} \mathbf{E}_i(R_{sp}\hat{\boldsymbol{n}},\omega)\mathbf{E}_j(R_{sp}\hat{\boldsymbol{n}},\nu) =\nonumber\\
&=\frac{1}{64\pi^{4}\varepsilon_{0}c^{3}}\int_{-\infty}^{+\infty}d\omega\,\int_{-\infty}^{+\infty}d\nu\,e^{i(\omega+\nu)(t-R_{sp}/c)}\sum_{i,j}q_{i}q_{j}J_{ij}(\omega,\nu)
\end{align}
where 
\begin{align}
J_{ij}(\omega,\nu)&:=\int d\Omega e^{i\left(\omega\bar{\boldsymbol{r}}_{i}+\nu\bar{\boldsymbol{r}}_{j}\right)\cdot\frac{\hat{\boldsymbol{n}}}{c}}\left[\ddot{\boldsymbol{r}}_{i}(\omega)\cdot\ddot{\boldsymbol{r}}_{j}(\nu)-\left(\ddot{\boldsymbol{r}}_{i}(\omega)\cdot\hat{\boldsymbol{n}}\right)\left(\ddot{\boldsymbol{r}}_{j}(\nu)\cdot\hat{\boldsymbol{n}}\right)\right].
\label{Jij}
\end{align}
The conditions for having coherent or incoherent emission, depending on the ratio between the system's dimension and the wavelength of the emitted radiation will appear clearly in the evaluation of $J_{ij}(\omega,\nu)$.  A rather long but straightforward calculation gives:
\begin{align}
J_{ij}(\omega,\nu)&=4\pi\ddot{\boldsymbol{r}}_{i}(\omega)\cdot\ddot{\boldsymbol{r}}_{j}(\nu)\frac{\left(b^{2}-1\right)\sin(b)+b\cos(b)}{b^{3}}-4\pi\ddot{\boldsymbol{r}}_{i}^{z}(\omega)\ddot{\boldsymbol{r}}_{j}^{z}(\nu)\frac{\left(b^{2}-3\right)\sin(b)+3b\cos(b)}{b^{3}}
\label{Jij2}
\end{align}
with $b=\frac{1}{c}|\omega\bar{\boldsymbol{r}}_{i}+\nu\bar{\boldsymbol{r}}_{j}|$. 

According to the CSL model, each particle's acceleration is given by
\begin{equation}
\ddot{\boldsymbol{r}}_{j}(t)=-\frac{\nabla_{j}V(\boldsymbol{r}_{1}(t),...,\boldsymbol{r}_{N}(t),t)}{m_{j}}
\end{equation}
with $V(\boldsymbol{r}_{1},...,\boldsymbol{r}_{N},t)$ introduced in Eq. (\ref{potential_new}). The corresponding Fourier transform is 
\begin{align}\label{accwapp}
\ddot{r}_{j}^{i}(\omega)&=\int dt\,e^{-i\omega t}\ddot{r}_{j}^{i}(t)=\frac{\hbar\sqrt{\lambda}}{m_{0}m_{j}}\int d\boldsymbol{r}\int dt\,e^{-i\omega t}\frac{\partial}{\partial r_{j}^{i}(t)}[\mu_{j}(\boldsymbol{r}_{j}(t)-\boldsymbol{r})]w(\boldsymbol{r},t)
\end{align}
where the lower index $j$ labels the different particles while the higher index $i$ labels the three spatial directions $x,\,y,\,z$.

Using again the rigid body assumption $\boldsymbol{r}_{j}(t)\simeq \bar{\boldsymbol{r}}_{j}$ and then introducing $\boldsymbol{s}=\bar{\boldsymbol{r}}_{j}-\boldsymbol{r}$ and $\tilde{w}(\boldsymbol{r},\omega):=\int dt\,e^{-i\omega t}w(\boldsymbol{r},t)$ one gets:
\begin{align}
\ddot{r}_{j}^{k}(\omega)=-\frac{\hbar\sqrt{\lambda}}{m_{0}m_{j}}\int d\boldsymbol{s}\left(\frac{\partial}{\partial s^{k}}\mu_{j}(\boldsymbol{s})\right)\tilde{w}(\bar{\boldsymbol{r}}_{j}-\boldsymbol{s},\omega).
\end{align}
Using the fact that the noise correlation in Eq. (\ref{noiscor}), in the frequency domain, implies
\begin{equation}
\mathbb{E}[\tilde{w}(\boldsymbol{r},\omega)\tilde{w}(\boldsymbol{r}',\nu)]=2\pi\delta(\omega+\nu) e^{-\frac{(\boldsymbol{r}-\boldsymbol{r}')^{2}}{4r_{C}^{2}}},
\end{equation}
one can compute:
\begin{equation}\label{Err}
\mathbb{E}[\ddot{\boldsymbol{r}}_{i}(\omega)\cdot\ddot{\boldsymbol{r}}_{j}(\nu)]=\frac{2\pi\hbar^{2}\lambda}{m_{0}^{2}m_{i}m_{j}}\delta(\omega+\nu)f_{ij}(\mu),
\end{equation}
where we introduced 
\begin{equation}
f_{ij}(\mu):=\sum_{k=x,y,z}f_{ij}^{k}(\mu) 
\end{equation}
and 
\begin{equation}
f_{ij}^{k}(\mu):=\int d\boldsymbol{s}\int d\boldsymbol{s}'e^{-\frac{(\bar{\boldsymbol{r}}_{i}-\bar{\boldsymbol{r}}_{j}+\boldsymbol{s}'-\boldsymbol{s})^{2}}{4r_{C}^{2}}}\left(\frac{\partial\mu_{i}(\boldsymbol{s})}{\partial s^{k}}\right)\left(\frac{\partial\mu_{j}(\boldsymbol{s}')}{\partial s'^{k}}\right).
\end{equation}

Clearly, exactly the same conclusions are true for the correlation $\mathbb{E}\left[\ddot{r}_{i}^{z}(\omega)\ddot{r}_{j}^{z}(\nu)\right]$, with the only difference of having $f_{ij}^{z}(\mu)$ instead of $f_{ij}(\mu)$ in Eq. (\ref{Err}).

We can finally go back to Eq. (\ref{Jij2}) for $J_{ij}(\omega,\nu)$, whose average gives:
\begin{align}
\mathbb{E}[J_{ij}(\omega,\nu)]&=\frac{8\pi^{2}\hbar^{2}\lambda}{m_{0}^{2}m_{i}m_{j}}\left[f_{ij}(\mu)\frac{\left(b^{2}-1\right)\sin(b)+b\cos(b)}{b^{3}}-f_{ij}^{z}(\mu)\frac{\left(b^{2}-3\right)\sin(b)+3b\cos(b)}{b^{3}}\right]\delta(\omega+\nu),
\end{align}
which simplifies significantly under the assumption $f_{ij}^{z}(\mu)=f_{ij}(\mu)/3$ (always fulfilled for spherical symmetric mass distributions which we will consider here):
\begin{equation}
\mathbb{E}[J_{ij}(\omega,\nu)]=\frac{8\pi^{2}\hbar^{2}\lambda}{m_{0}^{2}m_{i}m_{j}}\delta(\omega+\nu)f_{ij}(\mu)\left[\frac{2}{3}\frac{\sin(b)}{b}\right] 
\end{equation}
where we remind that $b=b_{ij}(\omega,\nu)=\frac{1}{c}|\omega\bar{\boldsymbol{r}}_{i}+\nu\bar{\boldsymbol{r}}_{j}|$. 
Then, the (noise averaged) power emission formula is 
\begin{align}\label{Ptot3}
\mathbb{E}\left[P(t)\right]&=\frac{1}{64\pi^{4}\varepsilon_{0}c^{3}}\int_{-\infty}^{+\infty}d\omega\,\int_{-\infty}^{+\infty}d\nu\,e^{i(\omega+\nu)(t-R_{sp}/c)}\sum_{i,j}q_{i}q_{j}\mathbb{E}\left[J_{ij}(\omega,\nu)\right]=\nonumber\\
&=\frac{\hbar^{2}\lambda}{12\pi^{2}\varepsilon_{0}c^{3}m_{0}^{2}}\!\int_{-\infty}^{+\infty}\!\!\!\!\!\!\!d\omega\,\sum_{i,j}\frac{q_{i}q_{j}}{m_{i}m_{j}}f_{ij}(\mu)\frac{\sin[b_{ij}(\omega,-\omega)]}{b_{ij}(\omega,-\omega)},
\end{align}
where $b_{ij}(\omega,-\omega)=\frac{\omega}{c}|\bar{\boldsymbol{r}}_{i}-\bar{\boldsymbol{r}}_{j}|$.
Using the relation 
\begin{equation}
P(t) = \int_{0}^{\infty} d\omega\,\hbar\omega\frac{d\Gamma_t}{d\omega}
\label{eq:P2}
\end{equation}
between the power emitted and the emission rate we obtain:
\begin{equation}
\frac{d\Gamma_{t}}{d\omega}=\frac{\hbar\lambda}{6\pi^{2}\varepsilon_{0}c^{3}m_{0}^{2}\omega}\,\sum_{i,j}\frac{q_{i}q_{j}}{m_{i}m_{j}}f_{ij}(\mu)\frac{\sin[b_{ij}(\omega,-\omega)]}{b_{ij}(\omega,-\omega)}
\label{rategen}
\end{equation}
Note that the factor ``6'' in the denominator Eq. (\ref{rategen}) in place of the factor ``12'' in Eq. (\ref{Ptot3}) is due to the fact that the extremes of integration in the variable $\omega$ in Eqs. (\ref{eq:P2}) and (\ref{Ptot3}) are not the same.

There are two relevant cases to be considered.

\noindent A) The average distance $|\bar{\boldsymbol{r}}_{i}-\bar{\boldsymbol{r}}_{j}|$ between the particles is much larger than the emitted wavelength  $\lambda_k=2\pi c/\omega$. Then, for $i \neq j$, one has $b_{ij}(\omega,-\omega)\gg1$ which implies:
\begin{equation}
\frac{\sin\left[b_{ij}(\omega,-\omega)\right]}{b_{ij}(\omega,-\omega)}\sim 0,
\end{equation}
i.e. the contribution from all these terms is negligible. Instead for $i=j$ one has:
\begin{equation}\label{limco}
\frac{\sin\left[b_{ij}(\omega,-\omega)\right]}{b_{ij}(\omega,-\omega)}= 1.
\end{equation} 
Therefore, only the terms with $i=j$ in the sum in Eq. (\ref{rategen}) survive, the particles emit incoherently and the emission rate is:
\begin{equation}
\frac{d\Gamma_{t}}{d\omega}=\frac{\hbar\lambda}{6\pi^{2}\varepsilon_{0}c^{3}m_{0}^{2}\omega}\,\sum_{i}\frac{q_{i}^{2}}{m_{i}^{2}}f_{ii}(\mu).
\label{rateinc}
\end{equation}
Using $f_{ii}(\mu)=3m_i^2/2r_C^2$ (valid for point-like particles, see Eqs. (\ref{enou}) and (\ref{fij}) below), dividing both sides of the equation by $\hbar$ and using $E=\hbar \omega$ we get 
\begin{equation}
\frac{d\Gamma_{t}}{dE}=\frac{\hbar\lambda}{4\pi^{2}\varepsilon_{0}c^{3}m_{0}^{2}r_C^2 E}\,\sum_{i}q_{i}^{2}.
\label{rateinc2}
\end{equation}
In particular, when all the charges and masses are the same, this emission rate grows \textit{linearly} with the number $N$ of particles of the system.

\noindent B) The average distance $|\bar{\boldsymbol{r}}_{i}-\bar{\boldsymbol{r}}_{j}|$ between the particles  is much smaller than the  emitted wavelength $\lambda$. Then for all $i,\, j$, one has $b_{ij}(\omega,-\omega)\ll1$ which implies Eq. (\ref{limco}) is valid, with good approximation, for all  terms. Accordingly, we have:
\begin{equation}
\frac{d\Gamma_{t}}{d\omega}=\frac{\hbar\lambda}{6\pi^{2}\varepsilon_{0}c^{3}m_{0}^{2}\omega}\,\sum_{i,j}\frac{q_{i}q_{j}}{m_{i}m_{j}}f_{ij}(\mu),
\label{ratecoh}
\end{equation}
here we must consider two sub-cases: 

\noindent B1) $r_{C}\gg|\bar{\boldsymbol{r}}_{i}-\bar{\boldsymbol{r}}_{j}|$. In this case we have
\begin{align}\label{enou}
f_{ij}^{k}(\mu)&\simeq\int d\boldsymbol{s}\int d\boldsymbol{s}'e^{-\frac{(\boldsymbol{s}'-\boldsymbol{s})^{2}}{4r_{C}^{2}}}\left(\frac{\partial\mu_{i}(\boldsymbol{s})}{\partial s^{k}}\right)\left(\frac{\partial\mu_{j}(\boldsymbol{s}')}{\partial s'^{k}}\right)=\int d\boldsymbol{s}\int d\boldsymbol{s}'\mu_{i}(\boldsymbol{s})\mu_{j}(\boldsymbol{s}')\left(\frac{\partial^{2}}{\partial s^{k}\partial s'^{k}}e^{-\frac{(\boldsymbol{s}'-\boldsymbol{s})^{2}}{4r_{C}^{2}}}\right)\nonumber\\
&=\int d\boldsymbol{s}\int d\boldsymbol{s}'\mu_{i}(\boldsymbol{s})\mu_{j}(\boldsymbol{s}')\frac{e^{-\frac{(\boldsymbol{s}'-\boldsymbol{s})^{2}}{4r_{C}^{2}}}}{2r_{C}^{2}}\left(1-\frac{(s'^{k}-s^{k})^{2}}{2r_{C}^{2}}\right).
\end{align}
Clearly this equality is exact when $i=j$. Assuming point-like mass densities i.e. $\mu_{i}(\boldsymbol{r})=m_{i}\delta(\boldsymbol{r})$ one gets $f_{ij}^{k}(\mu)=\frac{m_{i}m_{j}}{2r_{C}^{2}}$, which implies
\begin{equation}\label{fij}
f_{ij}(\mu)=\frac{3 m_{i}m_{j}}{2r_{C}^{2}}.
\end{equation}
The same conclusion is true also for particles with finite radious as long as this is much smaller than $r_C$, as it is for protons in the atomic nuclei. Replacing these results in Eq. (\ref{ratecoh}) and again dividing both sides by $\hbar$ and using $E=\hbar \omega$ we get
\begin{equation}
\frac{d\Gamma_{t}}{d E}=\frac{\hbar\lambda}{4\pi^{2}\varepsilon_{0}c^{3}m_{0}^{2}r_C^2 E}\,\left(\sum_{i} q_{i}\right)^2,
\label{ratecohreal}
\end{equation}
which describes coherent emission. In particular, when all the charges are the same, the emission rate of the system grows \textit{quadratically} with the number $N$ of particles.

\noindent B2) $r_C\ll|\bar{\boldsymbol{r}}_{i}-\bar{\boldsymbol{r}}_{j}|$. In this second case:
\begin{equation}
f_{ij}^{k}(\mu)\simeq e^{-\frac{(\bar{\boldsymbol{r}}_{i}-\bar{\boldsymbol{r}}_{j})^{2}}{4r_{C}^{2}}}\int d\boldsymbol{s}\int d\boldsymbol{s}'\left(\frac{\partial\mu_{i}(\boldsymbol{s})}{\partial s^{k}}\right)\left(\frac{\partial\mu_{j}(\boldsymbol{s}')}{\partial s'^{k}}\right)=0.
\end{equation}
where in the last step we used the fact that the mass density goes to zero at infinity. 
Therefore, in this regime the contribution of the mixed terms due to different particles is negligible i.e. the emission is incoherent and is given by Eq. (\ref{rateinc2}).

To summarize, in order to have coherent emission as described in Eq. (\ref{ratecohreal}), one needs to fulfill  two conditions:  ($i$) The average distance between the particles must be smaller than the wavelength of the emitted photons \textit{and} ($ii$) it must be smaller than $r_C$ . If one of these two conditions is not fulfilled, incoherent emission occurs as described in Eq. (\ref{rateinc2}).

\end{document}